# Nanoscale Electronic Inhomogeneity in $In_2Se_3$ Nanoribbons Revealed by Microwave Impedance Microscopy

*Keji Lai[1], Hailin Peng[2], Worasom Kundhikanjana[1], David T. Schoen[2], Chong Xie[2], Stefan Meister[2], Yi Cui[2], Michael A. Kelly[2], Zhi-Xun Shen[1]*

[1]Department of Applied Physics, Stanford University, Stanford, CA 94305

[2]Department of Materials Science and Engineering, Stanford University, Stanford, CA 94305

**Abstract**

Driven by interactions due to the charge, spin, orbital, and lattice degrees of freedom, nanoscale inhomogeneity has emerged as a new theme for materials with novel properties near multiphase boundaries. As vividly demonstrated in complex metal oxides[1-5] and chalcogenides[6,7], these microscopic phases are of great scientific and technological importance for research in high-temperature superconductors[1,2], colossal magnetoresistance effect[4], phase-change memories[5,6], and domain switching operations[7-9]. Direct imaging on dielectric properties of these local phases, however, presents a big challenge for existing scanning probe techniques. Here, we report the observation of electronic inhomogeneity in indium selenide ($In_2Se_3$) nanoribbons[10] by near-field scanning microwave impedance microscopy[11-13]. Multiple phases with local resistivity spanning six orders of magnitude are identified as the coexistence of superlattice, simple hexagonal lattice and amorphous structures with ~100nm inhomogeneous length scale, consistent with high-resolution transmission electron microscope studies. The atomic-force-microscope-compatible microwave probe is able to perform quantitative sub-surface electronic study in a noninvasive manner. Finally, the phase change memory function in $In_2Se_3$ nanoribbon devices can be locally recorded with big signal of opposite signs.

While the conventional wisdom on solids largely results from the real-space periodic structures and k-space band theories[14], recent advances in physics have shown clear evidence that microscopic inhomogeneity, manifested as sub-micron spatial variations of the material properties, could indeed occur under certain conditions. Utilizing various probe-sample coupling mechanisms[15], spatial inhomogeneity has been observed as nanometer gap variations in high-Tc superconductors[1,2], coexisting electronic states in $VO_2$ near the metal-insulator transition[3], ferromagnetic domains in manganites showing colossal magnetoresistance effect[4], and in an ever-growing list. Probing these non-uniform phases provides not only much knowledge of the underlying interactions, but also valuable information for applications of the domain structures. In particular, spatially resolved properties are of significant interest for phase change and other switching materials to be on board the nanoelectric era[5-9].

While a number of contrast mechanisms[15] have been employed to visualize the electronic inhomogeneity, established scanning probe techniques do not directly access the low-frequency (f) complex permittivity $\varepsilon(\omega) = \varepsilon' + i\sigma/\omega$, where $\varepsilon'$ is the dielectric constant and $\sigma$ the conductivity, which holds a special position to study the ground state properties of materials. For local electrodynamic response, near-field technique is imperative to resolve spatial variations at length scales well below the radiation wavelength[16]. For this study, the working frequency is set at ~1GHz, i.e., in the microwave regime, to stay below resonant excitations at optical frequencies[17]. The near-field interaction can be understood as tip-sample impedance change. And a thin surface insulating layer, equivalent to small series impedance, will not shield the capacitive coupling to sub-surface features. The AC detection is very sensitive using mostly noninvasive RF excitation (<0.1V), in contrast to electrostatic probes[15] requiring a high DC bias (>1V) to reveal the conductivity distribution.

The most common design of the near-field microwave probe is an etched metal tip, which limits the resolution to several microns, backed by a massive cavity or transmission line resonator[18-20]. The strong contact force due to the rigid structure easily causes damage to the tip and the sample, halting the applications on practical nano-devices. Compliant cantilever probes on atomic force microscope (AFM) platforms[21] (Fig. 1a) are much better in this aspect. In this work, the probe of our scanning microwave impedance microscope (MIM) is micro-fabricated on silicon nitride cantilevers with shielded metal traces and focused-ion beam (FIB) deposited Pt tip[11-13] (Fig.1a inset). The microwave electronics detect the real and imaginary components of the effective tip-sample impedance and output as MIM-R and MIM-C signals (Fig. 1b), which contain the local (ε,σ) information of the material. Therefore, we believe that the MIM can afford a powerful and general-purpose tool to study nanoscale dielectric inhomogeneity in a noninvasive manner.

The material for this study is the layer-structured III-VI compound $In_2Se_3$, a technologically important system in solar cell[22], battery[23], and phase-change random access memory (PRAM)[24,25] applications. The crystal structure, as seen in Fig. 1c, shows strong covalent bonding within each layer and weak van der Waals force between the layers. While bulk $In_2Se_3$ presents many complicated crystal structures[26], the nano-structures grown by the vapor-liquid-solid (VLS) method is single-crystalline in nature[27,28]. The nanoribbon samples, with a thickness $t$ = 20~50nm and a width varying from 100nm to several microns, were transferred to $SiN_x$ substrates and contacted by In/Au electrodes[10]. Fig. 1d shows the MIM image of an $In_2Se_3$ nanoribbon. While the standard AFM signal shows a flat surface, the MIM-C image reveals clear sub-micron spatial inhomogeneity. Throughout the paper, the color rendering is such that brighter color represents larger tip-sample capacitance in MIM-C, and more loss in MIM-R.

A careful study of the microwave images was performed on a large $In_2Se_3$ nanoribbon piece contacted by 4 In/Au electrodes, as seen from the scanning electron microscope (SEM) picture and AFM image in Fig. 2a. Voltage pulses were applied across these leads to induce appreciable conductivity inhomogeneity. After the pretreatment, the two-terminal resistances ($R_{2T}$)[29] of all three segments increased by 3 decades. Using the standard conductive AFM (C-AFM) mode, we also confirmed that the sample surface is insulating and prohibits DC current flow.

The MIM images of the same ribbon are shown in Fig. 2b, with the data on the metal electrodes excluded in the analysis. Substantial electronic inhomogeneity is observed within the nanoribbon, which cannot be resolved by SEM, AFM, or C-AFM. Qualitatively, the near-field response is consistent with the transport data between adjacent electrodes. For the upper two sections with $R_{2T}$ in the order of $10^8 \Omega$, the MIM images display strong non-uniformity. Whereas for the bottom section with $R_{2T}$ = 2MΩ, the bright area dominates the MIM-C image except for a weak breakage in the middle. The length scale of these inhomogeneous domains ranges from 100nm to microns. The sharpest transition between different domains is ~100nm, as seen in the line cut in Fig. 2c, which is equivalent to the tip diameter.

Quantitative understanding of the images requires finite-element analysis (FEA) of the tip-sample interaction[12] and the result is shown in Fig. 2c. For illustration purpose, the x-axis is converted to local sheet resistance $R_S = 1/(\sigma \cdot t)$ and y-axis to output voltage, with $1(G\Omega)^{-1}$ admittance (inverse impedance) contrast corresponding to ~1.5mV signal. For decreasing $R_S$, the MIM-C signal stays low in the resistive limit ($R_S > 10^8 \Omega$), rises monotonically in the crossover regime ($10^8 \Omega > R_S > 10^5 \Omega$), and finally saturates[30] in the conductive limit ($R_S < 10^5 \Omega$). The MIM-R signal, on the other hand, peaks around $R_S = 2M\Omega$, where the loss is significant, and vanishes at both extremes. The simulation is in excellent agreement with the data. For example, along the vertical line cut in Fig. 2b and 2c, the red dots mark the highly conductive regions, which are indeed bright in MIM-C and faint in MIM-R. The blue dots label the highly resistive areas with little contrast over the substrate in both channels. Two regions, bright in MIM-R but dim in MIM-C, are indicated by green dots. In particular, such intermediate-$\sigma$ feature traverses the entire ribbon between the bottom two electrodes, where $R_{2T} \sim 2M\Omega$ coincides with the crossover $R_S$ in Fig. 2c. Interestingly, the highly conductive regions are mostly found near the In/Au electrodes, which may indicate that the Joule heating first destroys the metallic phase away from the Ohmic contacts. Finally, the maximum contrast signal, ~300mV in MIM-C and ~100mV in MIM-R, also agrees with the lumped-element FEA simulation.

In order to shed some light on the observed electronic inhomogeneity, control experiments were carried out on $In_2Se_3$ nanoribbons deposited on $SiN_x$ membranes for simultaneous transmission electron microscope (TEM) and transport studies[10]. Three distinct structural phases, as seen from the high resolution TEM images and the selected-area electron diffraction (SAED) patterns in Fig. 3a, are observed in ribbons in the [11-20] growth direction. Correspondingly, typical I-V characteristics of the three phases are shown in Fig. 3b. It has been previously reported that pristine VLS-grown [11-20] $In_2Se_3$ nanoribbons with superlattice structures (Fig. 3a, left) exhibit metallic behavior and room temperature $R_{2T}$ as low as kilo-ohms[10]. Further studies show that after applying a low voltage (<1V) pulse, $R_{2T}$ can increase up to mega-ohms and the ribbons behave semiconducting. For these devices, the superlattice structure is usually lost and simple hexagonal lattice is observed (Fig. 3a, middle). Finally, as a phase change material[24,25], $In_2Se_3$ can be switched between crystalline and amorphous (Fig. 3a, right) states by high voltage pulses, with $R_{2T}$ in the giga-ohms range for the amorphous state. Interestingly, all three phases could coexist in a single device after Joule-heating, as seen in Fig. 3c and the zoom-in view in Fig. 3c. The length scale of this structural inhomogeneity is again around 50~100nm. We therefore assign the same color scheme (red, green, and blue) here in accordance with the local electrical imaging result in Fig. 2.

Finally, we demonstrate that the MIM can provide spatially resolved information when $In_2Se_3$ nano-devices are phase-switched by voltage pulses. Fig. 4a shows the SEM image of a long nanoribbon contacted by 9 In/Au leads. The resistance between adjacent fingers was around 10KΩ before any treatment. For the intact segment *iii* below the bottommost electrode, the MIM-C image in Fig. 4b (bottom) appears very bright over the substrate. In the successive transport measurements, short pulses, depicted in the inset of Fig. 4c, were applied to the ribbon and $R_{2T}$ increased to >1MΩ for all sections. After this pulse, the MIM-C image shows a much weaker contrast over the background, as seen for section *i* (Fig. 4b, top) with $R_{2T}$ = 2MΩ. Furthermore, a high bias sweep (Fig. 4c) was applied across the nanoribbon segment *ii*. Interestingly, the MIM-C data of this segment (Fig. 4b, middle) clearly break into two domains. The upper portion of the device shows positive contrast over the background and the lower portion negative. We therefore conclude that as the device cooled down from the melting point, the upper part re-crystallized and became highly conductive again, while the lower part stayed amorphous and resistive. We note that the large signal with opposite signs recorded by the MIM (Fig. 4c) most vividly displays the phase change memory function of the $In_2Se_3$ devices. It is possible to further implement the MIM as a spatially-resolved readout instrument for memories with large resistivity changes. Rather than detecting the minute topographical variations accompanying the phase change[6], the MIM directly measures the local electronic property and is much more sensitive for the operation.

**Reference:**

(1)   Pan, S. H.; O'Neal, J. P.; Badzey, R. L.; Chamon, C.; Ding, H.; Engelbrecht, J. R.; Wang, Z.; Eisaki, H.; Uchida, S.; Gupta, A. K.; Ng, K. W.; Hudson, E. W.; Lang, K. M.; Davis, J. C. *Nature* **2001**, *413*, 282-285.
(2)   McElroy, K.; Lee, J.; Slezak, J. A.; Lee, D. H.; Eisaki, H.; Uchida, S.; Davis, J. C. *Science* **2005**, *309*, 1048-1052.
(3)   Qazilbash, M. M.; Brehm, M.; Chae, B.-G.; Ho, P. C.; Andreev, G. O.; Kim, B.-J.; Yun, S. J.; Balatsky, A. V.; Maple, M. B.; Keilmann, F.; Kim, H.-T.; Basov, D. N. *Science* **2007**, *318*, 1750-1753.
(4)   Dagotto, E. *Nanoscale phase separation and colossal magnetoresistance: the physics of manganites and related compounds*; Springer: New York, 2002.

(5) Wu, J.; Gu, Q.; Guiton, B. S.; de Leon, N. P.; Ouyang, L.; Park, H. *Nano Letters* **2006**, *6*, 2313-2317.

(6) Hamann, H. F.; O'Boyle, M.; Martin, Y. C.; Rooks, M.; Wickramasinghe, H. K. *Nat. Mater.* **2006**, *5*, 383-387.

(7) Terabe, K.; Hasegawa, T.; Nakayama, T.; Aono, M. *Nature* **2005**, *433*, 47-50.

(8) Szot, K.; Speier, W.; Bihlmayer, G.; Waser, R. *Nat. Mater.* **2006**, *5*, 312-320.

(9) Jesse, S.; Rodriguez, B. J.; Choudhury, S.; Baddorf, A. P.; Vrejoiu, I.; Hesse, D.; Alexe, M.; Eliseev, E. A.; Morozovska, A. N.; Zhang, J.; Chen, L.-Q.; Kalinin, S. V. *Nat. Mater.* **2008**, *7*, 209-215.

(10) Peng, H.; Xie, C.; Schoen, D. T.; Cui, Y. *Nano Letters* **2008**, *8*, 1511-1516.

(11) Lai, K.; Ji, M. B.; Leindecker, N.; Kelly, M. A.; Shen, Z. X. *Review of Scientific Instruments* **2007**, *78*, 063702.

(12) Lai, K.; Kundhikanjana, W.; Kelly, M.; Shen, Z. X. *Review of Scientific Instruments* **2008**, *79*, 063703.

(13) Lai, K.; Kundhikanjana, W.; Kelly, M. A.; Shen, Z. X. *Appl. Phys. Lett.* **2008**, *93*, 123105.

(14) Ashcroft, N. W.; Mermin, N. D. *Solid State Physics*; Thomson Learning, 1976.

(15) Meyer, E.; Hug, H.; Bennewitz, R. *Scanning probe microscopy: the lab on a tip*; Springer: Berlin, 2003.

(16) Rosner, B. T.; van der Weide, D. W. *Review of Scientific Instruments* **2002**, *73*, 2505-2525.

(17) Hillenbrand, R.; Taubner, T.; Keilmann, F. *Nature* **2002**, *418*, 159-162.

(18) Wei, T.; Xiang, X.-D. *Appl. Phys . Lett.* **1996**, *68*, 3506-3508.

(19) Tabib-Azar, M.; Su, D. P.; Pohar, A.; LeClair, S. R.; Ponchak, G. *Review of Scientific Instruments* **1999**, *70*, 1725-1729.

(20) Alexander, T.; Steven, M. A.; Hans, M. C.; Robert, L. M.; Vladimir, V. T.; Andrew, R. S. *Review of Scientific Instruments* **2003**, *74*, 3167-3170.

(21) Massood, T.-A.; Yaqiang, W. *Microwave Theory and Techniques, IEEE Transactions on* **2004**, *52*, 971-979.

(22) Kwon, S. H.; Ahn, B. T.; Kim, S. K.; Yoon, K. H.; Song, J. *Thin Solid Films* **1998**, *323*, 265-269.

(23) Julien, C.; Hatzikraniotis, E.; Chevy, A.; Kambas, K. *Materials Research Bulletin* **1985**, *20*, 287-292.

(24) Lee, H.; Kang, D. H.; Tran, L. *Materials Science and Engineering B-Solid State Materials for Advanced Technology* **2005**, *119*, 196-201.

(25) Yu, B.; Ju, S. Y.; Sun, X. H.; Ng, G.; Nguyen, T. D.; Meyyappan, M.; Janes, D. B. *Appl. Phys. Lett.* **2007**, *91*, 133119.

(26) Ye, J. P.; Soeda, S.; Nakamura, Y.; Nittono, O. *Japanese Journal of Applied Physics Part 1-Regular Papers Short Notes & Review Papers* **1998**, *37*, 4264-4271.

(27) Peng, H. L.; Schoen, D. T.; Meister, S.; Zhang, X. F.; Cui, Y. *Journal of the American Chemical Society* **2007**, *129*, 34-35.

(28) Two different growth directions, [0001] and [11-20], are found in our VLS-grown $In_2Se_3$ nano-structures. For simplicity, we only discuss the results of nanoribbons with [11-20] growth direction, i.e., perpendicular to the normal direction of the layers.

(29) Only $R_{2T}$ is reported in the paper because the resistance of $In_2Se_3$ ribbons spans many orders of magnitude. For metallic ribbons, 4-terminal measurements were also performed and it is confirmed that the In/Au contact resistance is small (< 1 KΩ) compared with the sample resistance.

(30) Wang, Z.; Kelly, M. A.; Shen, Z.-X.; Shao, L.; Chu, W.-K.; Edwards, H. *Appl. Phys. Lett.* **2005**, *86*, 153118.

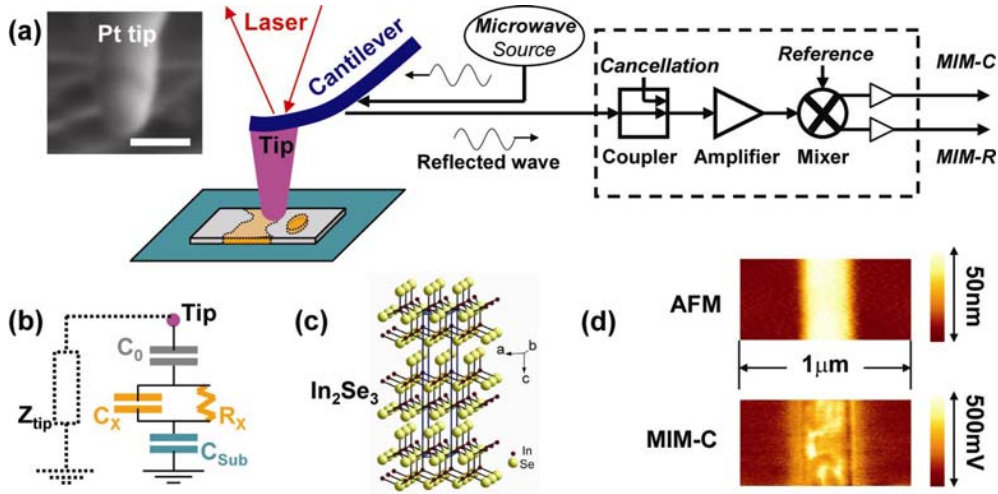

Figure 1. (a) Schematic of the cantilever-based MIM system setup. Surface topography is simultaneously obtained by the AFM laser feedback. The microwave electronics suppress the background of the reflected microwave signal and detect the changes during the scanning. The two mixer outputs are recorded to form MIM-C and MIM-R images. The inset SEM picture shows the Pt tip deposited by focused-ion beam (FIB), and the scale bar is 200nm. (b) A circuit model of the tip-sample interaction, shown as a lumped-element load in parallel with the tip impedance $Z_{tip}$. The surface layer and the substrate are both dielectrics and represented by capacitors $C_0$ and $C_{sub}$. The $In_2Se_3$ sample with finite conductivity is described as two components $C_X$ and $R_X$ in parallel. (c) Layered crystal structure of bulk $In_2Se_3$. (d) Simultaneously taken topography and MIM-C images of an $In_2Se_3$ nanoribbon. The resistance across this device is 2MΩ. The microwave image reveals clear electronic inhomogeneity, while the AFM image shows essentially a flat surface.

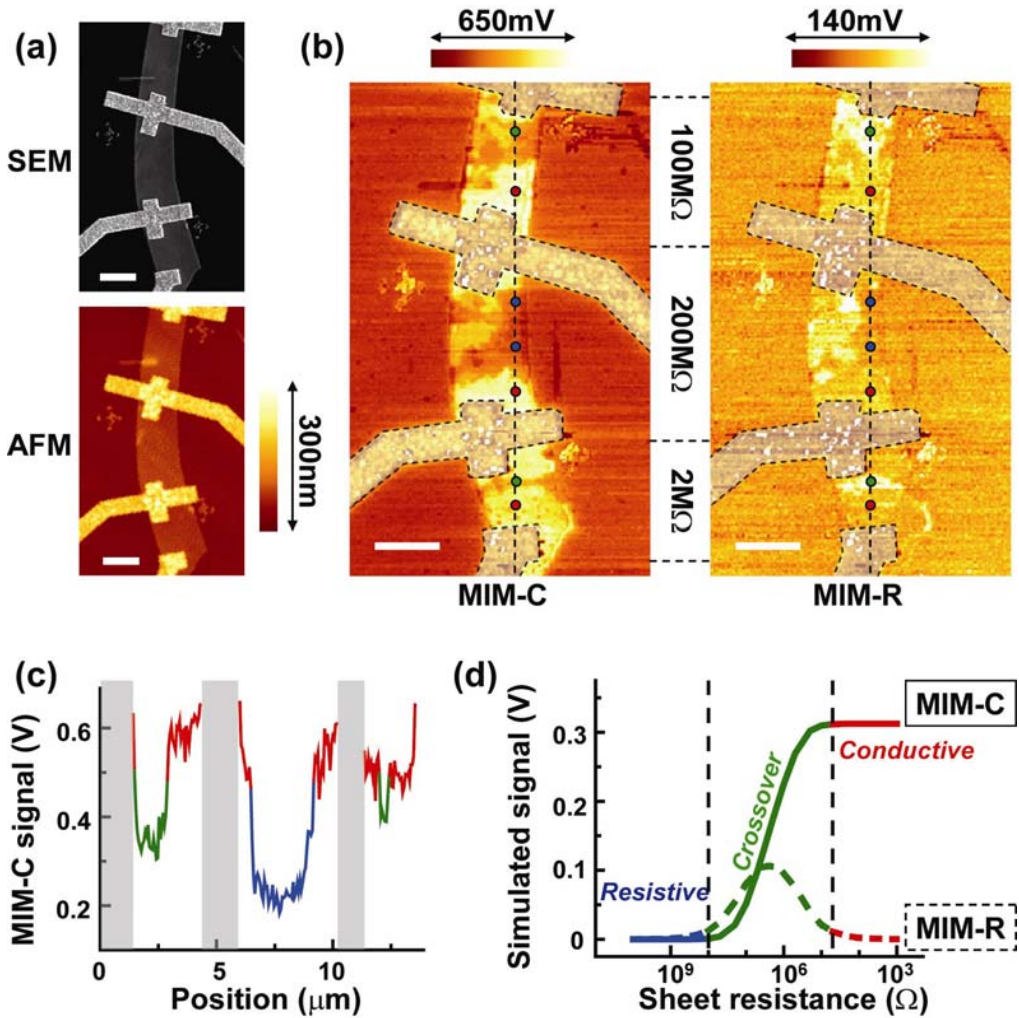

Figure 2. (a) SEM (top) and AFM (bottom) image of a large $In_2Se_3$ nanoribbon. (b) MIM-C (left) and MIM-R (right) images of the same ribbon, with the electrodes masked. The resistances of the three segments (top-down) are 100MΩ, 200MΩ, and 2MΩ, respectively. A dashed line across the ribbon is plotted and labeled by red, green, and blue dots. All scale bars in (a) and (b) are 2μm. (c) A line cut of the MIM-C image in (b) (electrodes excluded). (d) Simulated MIM-C (solid line) and MIM-R (dashed line) signal as a function of the local sheet resistance. The black dashed lines roughly mark the boundaries of different regimes. The MIM-C signal increases monotonically with reducing sheet resistance, while the MIM-R signal peaks around 2MΩ/square in the crossover region. (b), (c), and (d) are color coded such that red is used for highly conductive, green for intermediate and blue for highly resistive areas.

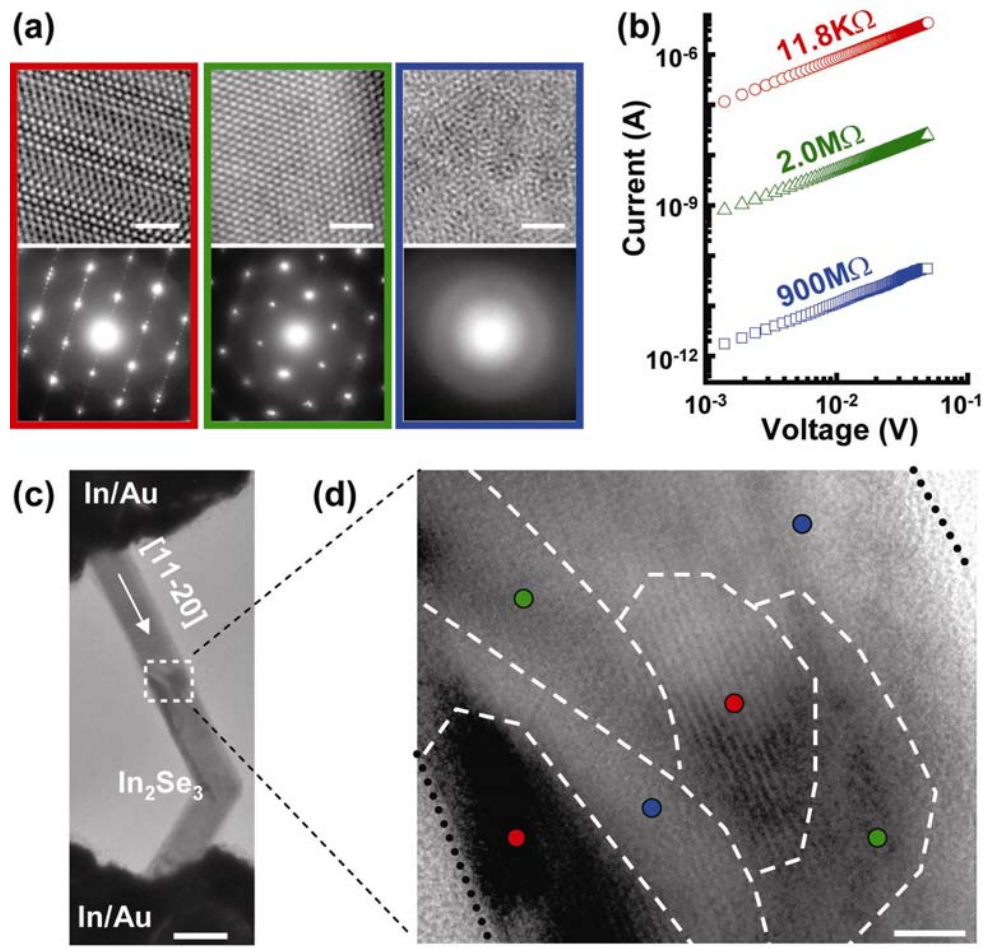

Figure 3. (a) High-resolution transmission electron microscopy (HRTEM) images (top) and the corresponding selected-area electron diffraction (SAED) patterns (bottom). The structures from left to right correspond to superlattice, simple hexagonal lattice, and amorphous phases. All scale bars are 2nm. (b) Typical I-V characteristics of three different states of [11-20] $In_2Se_3$ nanoribbons. (c) Low-resolution TEM image of an $In_2Se_3$ nanoribbon device on $SiN_x$ membrane. The scale bar is 100nm. (c) Zoom-in HRTEM image of the selected area in (c). The scale bar is 10nm. The black dotted lines show the boundaries of the ribbon. The dashed white lines roughly mark the boundaries between different domains. The same color scheme (red, green, and blue) as that in Figure 2 is used.

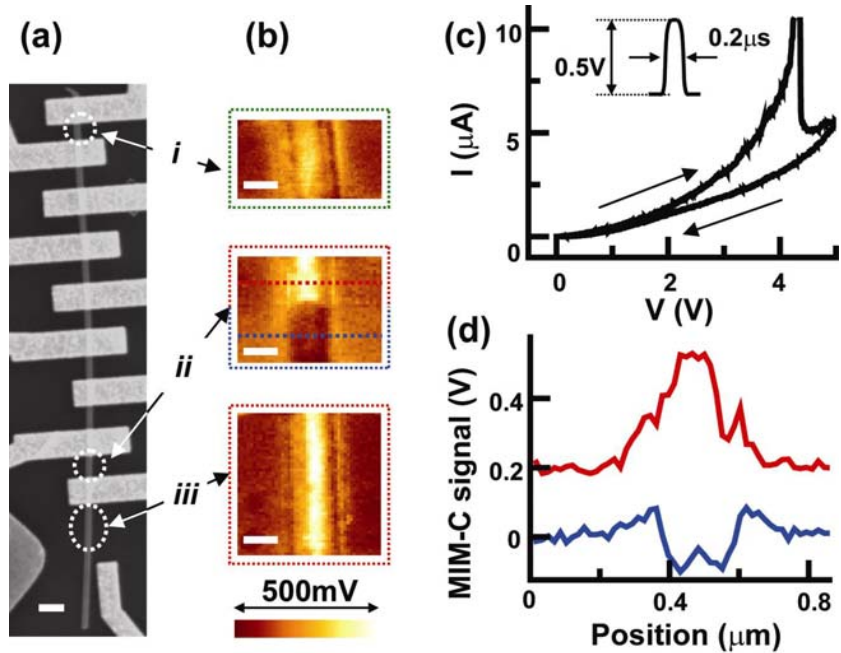

Figure 4. (a) SEM image of a long In$_2$Se$_3$ nanoribbon, 200nm in width and 30nm in thickness, with multiple contacts. The scale bar is 1μm. (b) MIM-C images of three segments, labeled as *i*, *ii*, and *iii* in (a), of the same ribbon. The scale bars are 200nm. (c) I-V curve of a high voltage sweep on segment *ii*. For the up-sweep, the current increased above 10μA at around 4V bias and suddenly dropped to 5μA. The down-sweep then followed completely different characteristics. The inset shows a low voltage (0.5V) pulse applied on the nanoribbon before this sweep. (c) Two line cuts of the MIM-C image ((b)) of segment *ii*, showing two domains with opposite contrast over the background. The red line is shifted upward by 0.2V for clarity. Again, the color coding (red, green, and blue) is in accordance with that in Figures 2 and 3.